\documentclass[12pt]{article}
\usepackage{graphicx}

\newcommand{\Frac}[2]{\frac{\displaystyle #1}{\displaystyle #2}}
\newcommand{\beq}{\begin{equation}}
\newcommand{\eeq}{\end{equation}}
\newcommand{\beqn}{\begin{eqnarray}}
\newcommand{\eeqn}{\end{eqnarray}}
\newcommand{\beqns}{\begin{eqnarray*}}
\newcommand{\eeqns}{\end{eqnarray*}}

\begin{document}

\begin{titlepage}
\begin{center}

\hfill USTC-ICTS-12-17\\
\hfill November 2012

\vspace{2.5cm}

{\large {\bf  Angular distributions of
$\tau^-\to K_S\pi^-\nu_\tau$ decay}}\\
\vspace*{1.0cm}
 {Dao-Neng Gao$^\dagger$ and Xian-Fu Wang$^\ddagger$} \vspace*{0.3cm} \\
{\it\small Interdisciplinary Center for Theoretical Study,
University of Science and Technology of China, Hefei, Anhui 230026
China}

\vspace*{1cm}
\end{center}
\begin{abstract}
\noindent
 Based on experimental data by the Bell Collaboration, we present a phenomenological analysis of the angular distributions in $\tau^-\to K_S\pi^-\nu_\tau$ decay. Our study shows that the angular analysis could lead to some interesting observables, and the future experimental investigation of these observables might be very helpful in revealing the nature of the scalar components of the decay. New physics contributions from the two-Higgs-doublet model to this decay have also been examined.
\end{abstract}

\vfill
\noindent
$^{\dagger}$ E-mail:~gaodn@ustc.edu.cn\\\noindent
$^{\ddagger}$ E-mail:~wangxf5@mail.ustc.edu.cn
\end{titlepage}

\section{Introduction}
The $\tau$ lepton is the only known lepton massive enough to decay
into hadrons and its hadronic decays provide a very useful laboratory to test the low energy
dynamics of the standard model (SM) \cite{Pich97, DHZ06}.
With the increased experimental
sensitivities achieved already or in the future, some interesting
limits on possible new physics contributions to the $\tau$ decay
amplitudes could be expected; for instance, in the decays $\tau^\pm \to K_S \pi^\pm \nu_\tau$,
upper limits on the CP violation parameter from multi-Higgs-doublet models have been obtained
by Belle Collaboration \cite{Belle2011}.

The decay $\tau \to K\pi \nu_\tau$, which has the largest branching ratio of all Cabibbo-suppressed decays, could be a powerful probe of the strange sector of the weak charged current \cite{KM92, DM93, FM96}. Early investigations have established that the dominant contribution to the decay spectrum is from the $K^*(892)$ meson \cite{ExpKstar}, while the scalar or tensor contributions, which are expected theoretically \cite{FM96, NC95}, are not excluded experimentally \cite{Aleph99, Cleo96}. Recently, a precise measurement of $\tau^-\to K_S\pi^-\nu_\tau$ decay has been done by Belle Collaboration \cite{Belle2007} with high-statistics data, which showed that the $K^*(892)$ alone is not sufficient to describe the $K_S\pi$ invariant mass spectrum and other states such as scalar ($K^*_0(800)$, $K^*_0(1430)$, ...) or more vector resonances ($K^*(1410)$, $K^*(1680)$, ...) have to be included. Thus three models $K^*_0(800)+K^*(892)+K^*(1410)$, $K^*_0(800)+K^*(892)+K^*_0(1430)$, and $K^*_0(800)+K^*(892)+K^*(1680)$ were introduced to fit the data, and the best description of the decay spectrum is achieved in the first two models \cite{Belle2007}. One can expect that these two models will lead to the different angular distribution behavior since they have different scalar and vector components. This angular analysis has not been done yet experimentally.

Motivated by the above new experimental data, in this paper we present a phenomenological analysis of the angular distributions in the $\tau^-\to K_S\pi^- \nu_\tau$ decay.
We hope that this angular analysis will reveal the difference from the above models, which could further increase our understanding of the scalar contributions in this decay.

The paper is organized as follows. In section 2, we will discuss the decay distributions of $\tau^-\to K_S\pi^-\nu_\tau$, and some interesting observables will be introduced. In Section 3, a phenomenological analysis of the angular distribution is carried out, and new physics contributions  from two-Higgs-doublet model to this decay will be examined. Finally, we summarize our results in Section 4.

\section{Decay distributions}

In the SM, the general invariant amplitude of $\tau^-\to K_S\pi^-\nu_\tau$ can be decomposed as a product of a leptonic current and a hadronic current \cite{FM96}
\beq\label{invariantM1}{\cal M}=\frac{G_F\sin\theta_C}{\sqrt{2}}M_\mu J^\mu. \eeq
Here $G_F$ is the Fermi coupling constant and $\theta_C$ is the Cabibbo angle. The leptonic current is given by
\beq\label{leptonic}
M_\mu=\bar{u}(p_{\nu_\tau})\gamma_\mu(1-\gamma_5)u(p_\tau),
\eeq
and the hadronic current can be parameterized by two form factors as
\beq\label{hadronic}
J^\mu=\langle K_S\pi^-|\bar{s}\gamma^\mu u|0\rangle =F_V(s)\left(g^{\mu\nu}-\frac{Q^\mu Q^\nu}{Q^2}\right)q_\nu+F_S(s)Q^\mu \eeq
with $$ Q^\mu=p_K^\mu+p_\pi^\mu, \qquad q_\mu=p_K^\mu-p_\pi^\mu, \qquad s=Q^2,$$
where $F_V$ is the vector form factor, corresponding to the $J^P=1^-$ component of the strange weak charged currents, and $F_S$ is the scalar one, corresponding to the $J^P=0^+$ component. In the limit of SU(3) symmetry, $m_K^2=m_\pi^2$, the vector current is conserved and $F_S$ is zero. Now the differential decay rate, in terms of $s$, the $K_S \pi$ invariant mass squared, and $\theta$, the angle between the three-momentum of $K_S$ and the three momentum of $\tau^-$ in the $K_S\pi^-$ rest frame, can be written as
\beqn\label{doubleDG}
\frac{d^2\Gamma}{ds ~d\cos\theta}&=&\frac{G_F^2\sin^2\theta_C}{2^6\pi^3\sqrt{s}}\frac{(m_\tau^2-s)^2}{m_\tau^3}P(s)\left[\left(\frac{m_\tau^2}{s}\cos^2\theta+
\sin^2\theta\right)P^2(s) |F_V|^2\right.\nonumber\\
&&\left. +\frac{m_\tau^2}{4}|F_S|^2-\frac{m_\tau^2}{\sqrt{s}} P(s) {\rm Re}(F_V F_S^*)\cos\theta\right]
\eeqn
with \[P(s)=\frac{1}{2\sqrt{s}}\sqrt{(s+m_K^2-m_\pi^2)^2-4 s m_K^2} ,\]
and the
phase space is given by \beq\label{phasespace} (m_\pi+m_K)^2\le s\le m_\tau^2,\;\;\;\;
-1\le\cos\theta\le 1 .\eeq
After integrating over $\cos\theta$ in eq. (\ref{doubleDG}), one can get
\beq\label{dgammaoverds}\frac{d\Gamma}{ds}=\frac{G_F^2\sin^2\theta_C}{2^6\pi^3\sqrt{s}}\frac{(m_\tau^2-s)^2}{m_\tau^3}P(s)\left[\left(\frac{2m_\tau^2}{3s}
+\frac{4}{3}\right)P^2(s) |F_V|^2+\frac{m_\tau^2}{2}|F_S|^2\right].\eeq
It is generally believed that this decay spectrum is dominated by the vector contributions ($K^*(892)$ resonances) \cite{ExpKstar}. Recent experimental results from the Bell Collaboration \cite{Belle2007} have shown that the scalar contribution is necessary to fit the data, although it provides only a small contribution to the decay rate.  It is found in \cite{Belle2007} that two different combinations  $K^*_0(800)+K^*(892)+K^*(1410)$ and $K^*_0(800)+K^*(892)+K^*_0(1430)$ can both describe the spectrum very well.  In order to further understand the scalar form factor, one may carry out the angular analysis.  From eq. (\ref{doubleDG}), if we only focus on the angular part, we can write the distribution as
\beq\label{angulardis1}\frac{d\Gamma}{d\cos\theta}=I_0+ I_1 \cos\theta+I_2 \cos^2\theta,
\eeq
with
\beqn\label{I0}&&I_0=\int^{s_{\rm max}}_{s_{\rm min}}ds\frac{G_F^2\sin^2\theta_C}{2^6\pi^3\sqrt{s}}\frac{(m_\tau^2-s)^2}{m_\tau^3}P(s)\left(\frac{m_\tau^2}{4}|F_S|^2+P^2(s)|F_V|^2\right),\\\nonumber\\
\label{I1}&&I_1=\int^{s_{\rm max}}_{s_{\rm min}}ds\frac{G_F^2\sin^2\theta_C}{2^6\pi^3\sqrt{s}}\frac{(m_\tau^2-s)^2}{m_\tau^3}P^2(s)\frac{-m_\tau^2}{\sqrt{s}}{\rm Re}(F_V F_S^*),\\\nonumber\\
\label{I2}&&
I_2=\int^{s_{\rm max}}_{s_{\rm min}}ds\frac{G_F^2\sin^2\theta_C}{2^6\pi^3\sqrt{s}}\frac{(m_\tau^2-s)^2}{m_\tau^3}P^3(s)\frac{m_\tau^2-s}{s}|F_V|^2.
\eeqn
Here $(s_{\rm min}, s_{\rm max})$ denotes the range of the integration over $s$, which could be the full phase space shown in eq. (\ref{phasespace}) or some kinematical cuts on $s$.  Using the above equations, one can further get
\beq\label{angulardis2} \frac{1}{\Gamma}\frac{d\Gamma}{d\cos\theta}={\cal G}+{\cal A}\cos\theta+\frac{3}{2}(1- 2{\cal G})\cos^2\theta,\eeq
with \beq\label{decayrate}\Gamma=2(I_0+\frac{1}{3}I_2)\eeq
is the decay rate. The constant term in (\ref{angulardis2}) can be expressed as \beq\label{gterm}{\cal G}=\frac{I_0}{\Gamma},\eeq
and the linear term in $\cos\theta$ \beq\label{asymmetry}{\cal A}=\frac{I_1}{\Gamma}\eeq
will give a vanishing contribution to the decay rate after integrating $\cos\theta$ in the full phase space. However, this term can induce an interesting observable, called the forward-backward asymmetry. Eq. (\ref{asymmetry}) gives the integrated and normalized asymmetry.  One can also define the differential asymmetry as \cite{BT95, KLM12}\beq\label{asym1}
 {A}_{\rm FB}(s)=\Frac{\int^1_0 \left(\frac{d^2\Gamma}{ds~ d\cos\theta}\right)d\cos\theta-\int^0_{-1}
 \left(\frac{d^2\Gamma}{ds~ d\cos\theta}\right)d\cos\theta}{\int^1_0 \left(\frac{d^2\Gamma}
 {ds~ d\cos\theta}\right)d\cos\theta+\int^0_{-1}
 \left(\frac{d^2\Gamma}{ds~ d\cos\theta}\right)d\cos\theta}.\eeq
Thus together with eq. (\ref{doubleDG}), we have
\beq\label{asym2}
{A}_{\rm FB}(s)=\Frac{-\frac{P(s)}{\sqrt{s}} {\rm Re}(F_V F_S^*)}{\frac{2}{3s}\left(1+\frac{2s}{m_\tau^2}\right)P^2(s)|F_V|^2+\frac{1}{2}|F_S|^2}.
\eeq
Note that the forward-backward asymmetries are generated from the interference between the scalar part and vector part amplitudes of the decay. In the case that the scalar contribution is not large, study of these asymmetries may be very important for us to extract the information on the scalar form factor.

\section{Phenomenological analysis}

In order to evaluate the quantities ${\cal G}$, ${\cal A}$, and $A_{\rm FB}$ defined in the previous section, we need the information on the hadronic form factors $F_V$ and $F_S$.
Theoretically, due to the nonperturbative Quantum Chromodynamics (QCD) effects at low energies, there is no model independent way at the present.  Therefore, various methods have been employed to study these form factors of $\tau\to K\pi \nu_\tau$ decays, such as meson dominance models \cite{FM96, FZ99} and chiral lagrangian including the resonances \cite{JPP06, BEJ, KLMN09,KLM12}. Phenomenologically, as shown in \cite{Belle2007, FM96}, these form factors could be parameterized by resonances contributions, which read
\beq\label{FVBR}
F_V=\frac{1}{\sqrt{2}(1+\beta+\chi)}\left[{\rm BW}_{K^*(892)}(s)+\beta {\rm BW}_{K^*(1410)}(s)+\chi {\rm BW}_{K^*(1680)}(s)\right]
\eeq
and
\beq\label{FSBR}
F_S=\frac{1}{\sqrt{2}}(m_K^2-m_\pi^2)\left[\frac{\kappa}{M^2_{K^*_0(800)}}{\rm BW}_{K^*_0(800)}(s)+\frac{\gamma}{M^2_{K^*_0(1430)}}{\rm BW}_{K^*_0(1430)}(s)\right],
\eeq
where
\[{\rm BW}_R(s)=\frac{M^2_R}{s-M^2_R+i \sqrt{s} \Gamma_R(s)}\]
and
\[\Gamma_R(s)=\Gamma_{0R}\frac{M^2_R}{s}\left(\frac{P(s)}{P(M_R^2)}\right)^{2\ell+1}\]
is the $s$-dependent total width of the resonance. $\ell=1(0)$ for $K\pi$ in the $P(S)$-wave state and $\Gamma_{0R}$ is the width of the resonance at its peak. $\beta$, $\chi$, $\kappa$, and $\gamma$ in eqs.(\ref{FVBR}) and (\ref{FSBR}) are parameters for the fractions of the resonances, which can be determined by fitting the data of the hadronic invariant mass distribution.

Actually these parameters have been determined in Ref. \cite{Belle2007} for two models with different combinations of the resonances, both of which can provide a good description of the decay spectrum. For the model I: $K^*_0(800)+K^*(892)+K^*(1410)$,
\beqn\label{modelI} |\beta|=0.075\pm 0.006,\;\;\;{\rm arg}(\beta)=1.44\pm 0.15,\;\;\; \kappa=1.57\pm 0.23. \eeqn
For the model II: $K^*_0(800)+K^*(892)+K^*_0(1430)$, the parameters could have two solutions from the data fit, which read (hereafter we call them as Model II-1 and Model II-2, respectively)
\beqn
&&\texttt{Model~II-1}:\;\; \kappa=1.27\pm 0.22,\;\;\; |\gamma|=0.954\pm 0.081,\;\; {\rm arg}(\gamma)=0.62\pm0.34, \label{modelIIs1}\\
&&\texttt{Model~II-2}:\;\; \kappa=2.28\pm 0.47,\;\;\; |\gamma|=1.92\pm 0.20,\;\; {\rm arg}(\gamma)=4.03\pm0.09.\label{modelIIs2}
\eeqn
It is obvious that there is no contribution from $K^*(1680)$ in these two models, $\chi$ in eq.(\ref{FVBR}) is set to zero.

\begin{table}[t]\begin{center}\begin{tabular}{c| c c c} \hline\hline\\
 &~~ Model I~~~~~~& Model II-1~~~~&~ Model II-2~~
\\$s_{\rm min}\sim s_{\rm max}$& & & \\\hline\\
 $(m_K+m_\pi)^2\sim$(0.8 GeV)$^2$ & 0.326$\pm$0.021 ~~~  &  0.320$\pm$0.021 ~~ & 0.343$\pm$0.039
\\(0.8 GeV)$^2\sim$ (1.0 GeV)$^2$& 0.257$\pm$0.001 ~~~  & 0.257$\pm$0.001 ~~ & 0.257$\pm$0.002   \\
 (1.0 GeV)$^2\sim$(1.2 GeV)$^2$& 0.333$\pm$0.005 ~~~  & 0.340$\pm$0.007 ~~  & 0.339$\pm$0.008  \\
 (1.2 GeV)$^2\sim$(1.4 GeV)$^2$& 0.398$\pm$0.003 ~~~  & 0.436$\pm$0.006 ~~  &0.437$\pm$0.009\\
 (1.4 GeV)$^2\sim$$m_\tau^2$& 0.440$\pm$0.002 ~~~  & 0.479$\pm$0.002 ~~  &0.479$\pm$0.005\\
$(m_K+m_\pi)^2\sim m_\tau^2$ &0.269$\pm$0.002 ~~~ &0.270$\pm$0.003 ~~ & 0.271$\pm$0.004\\\hline\hline
\end{tabular}\caption{The values of ${\cal G}$ defined in eq. (\ref{gterm}) are evaluated for different cuts on $s$ in the above models [(\ref{modelI}), (\ref{modelIIs1}), and (\ref{modelIIs2})]. The last line is for the full phase space.}\end{center}\end{table}

\begin{table}[t]\begin{center}\begin{tabular}{c| c c c} \hline\hline\\
 &~~ Model I~~~~~~& Model II-1~~~~&~ Model II-2~~
\\$s_{\rm min}\sim s_{\rm max}$& & & \\\hline\\
 $(m_K+m_\pi)^2\sim$(0.8 GeV)$^2$ & -0.229$\pm$0.006 ~~  &  -0.300$\pm$0.030 ~~ & -0.217$\pm$0.022
\\(0.8 GeV)$^2\sim$ (1.0 GeV)$^2$& -0.107$\pm$0.015 ~~  & -0.108$\pm$0.016 ~~ & -0.105$\pm$0.032   \\
 (1.0 GeV)$^2\sim$(1.2 GeV)$^2$& -0.332$\pm$0.039 ~~  & -0.182$\pm$0.073 ~~  & -0.413$\pm$0.075  \\
 (1.2 GeV)$^2\sim$(1.4 GeV)$^2$& -0.378$\pm$0.044 ~~  & -0.292$\pm$0.116 ~~  &-0.042$\pm$0.110\\
 (1.4 GeV)$^2\sim$$m_\tau^2$& -0.297$\pm$0.037 ~~  & -0.481$\pm$0.044 ~~  &0.471$\pm$0.019\\
$(m_K+m_\pi)^2\sim m_\tau^2$ &-0.133$\pm$0.018 ~~ &-0.125$\pm$0.021 ~~ & -0.123$\pm$0.037\\\hline\hline
\end{tabular}\caption{The integrated and normalized forward-backward asymmetry ${\cal A}$ are evaluated for different cuts on $s$ in the above models [(\ref{modelI}), (\ref{modelIIs1}), and (\ref{modelIIs2})]. The last line is for the full phase space.}\end{center}\end{table}

Now we can calculate the quantity ${\cal G}$ and the forward-backward asymmetry ${\cal A}$. The results have been shown in Tables 1 and 2, respectively. It is found that, when these two quantities are evaluated for the full phase space, all the models give consistent results. Model II-1 and II-2 are consistent for ${\cal G}$ even if we calculate it for different cuts on $s$. However, in the large $s$ region, $s\geq$(1.2 GeV)$^2$, it is expected that one might distinguish Model I and Model II from the values of ${\cal G}$. The asymmetry ${\cal A}$ shown in Table 2 is more interesting since its values will show the difference from these three cases for $s$ above $1~{\rm GeV}^2$.  A similar conclusion can also be obtained from the differential asymmetry $A_{\rm FB}$, which has been plotted in Figure 1. $A_{\rm FB}$'s are almost consistent for three cases when $s$ is below 1 GeV$^2$; however, they behave differently for $s$ above $1~{\rm GeV}^2$, in particular, $A_{\rm FB}$ from Model II-2 could change sign for large $s$.

\begin{figure}[t]
\begin{center}
\includegraphics[width=12cm,height=10cm]{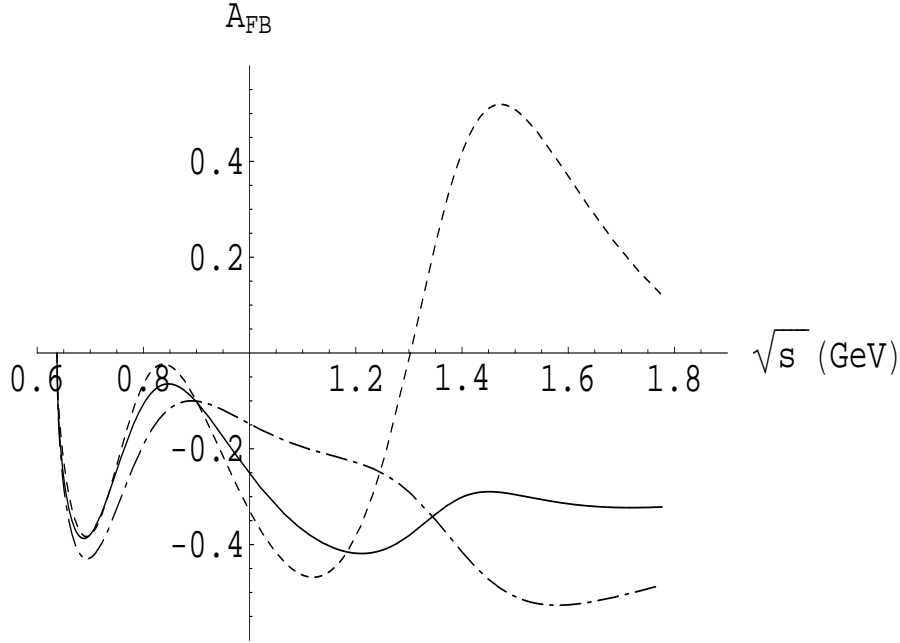}
\end{center}
\caption{The differential forward-backward asymmetry $A_{\rm FB}$ is plotted as the function of $\sqrt{s}$. The solid line is for Model I, the dashed-dotted line is for Model II-1, and the dashed line is for Model II-2.}
\end{figure}

Early studies tell us that the decay $\tau \to K\pi \nu_\tau$ is dominated by vector $K^*(892)$ contributions. The above different models [(\ref{modelI}), (\ref{modelIIs1}), and (\ref{modelIIs2})] from the Belle experiment have parameterized different scalar contents; therefore, it is  not surprising that analysis of the angular distribution of the decay will reveal some difference. This means that the future angular analysis from the high-statistics experiment might be useful in distinguishing the above models, and would help us to further understand the scalar components of the decay.

We would like to give some remarks here.
\begin{itemize}

\item The purpose of this paper is to show that the angular analysis of the decay $\tau^- \to K_S\pi^-\nu_\tau$ may help us to reveal the nature of the scalar form factors. In order to illustrate our analysis, we simply take the form factors adopted in the Belle experiment \cite{Belle2007}. It is not surprising that other different descriptions of the form factors, in particular, for the scalar form factors, could lead to different results from those in the present work.

\item Actually, some physically better motivated descriptions of the scalar form factors, which both satisfy constraints by analyticity and unitarity and provide a good description of the experimental data, have been obtained in a series of seminal papers \cite{JOP}. Note that our $F_S$ is equivalent to $(m_K^2-m_\pi^2)/{s} \cdot F^{K\pi}_0$ (here $F^{K\pi}_0$ is the scalar form factor) in these papers due to different notations. Thus the different behavior of the differential forward-backward asymmetry $A_{\rm FB}$ for the small $s$ region can be expected, to be compared with those shown in Fig. 1, in which $A_{\rm FB}$'s from the Belle data are consistent with one another for $s$ below 1 GeV$^2$. A detailed quantitative comparison is beyond the scope of the present paper, which is left for future work.

\item For comparison, we suggest our experimental colleagues should also include these physically better motivated form factors in the future experimental analysis.
\end{itemize}

In Ref. \cite{KLM12}, the differential forward-backward asymmetry $A_{\rm FB}$ has been studied in the two-Higgs-doublet model (2HDM) of type II with  large tan$\beta$ \cite{2HDM}. In this model, the charged Higgs exchange
\beq\label{tree} {\cal
L}^{H^\pm}=\frac{G_F}{\sqrt{2}}\sin\theta_C
\frac{m_s m_\tau\tan^2\beta}{m_{H^\pm}^2}\bar{\nu_\tau}(1+\gamma_5)\tau
\bar{s}(1-\gamma_5)u,\eeq
can lead to a tree level contribution to $\tau\to K\pi\nu_\tau$, which thus gives a new contribution to the scalar form factor $F_S$,
\beq\label{FSnew}
F_S^{\rm New}=F_S+\frac{1}{\sqrt{2}}\frac{m_K^2 \tan^2{\beta}}{M_{H^\pm}^2}\frac{m_s}{m_s+m_u}.
\eeq
In general, this new contribution will be strongly suppressed by the large charged Higgs mass $M_{H^\pm}$, which however may be substantially compensated by large tan$\beta$. In terms of $F_S^{\rm New}$, together with eq. (\ref{asym2}), one can evaluate the new contribution to $A_{\rm FB}$, which has been plotted in Figure 2. We take $F_S$ in eq. (\ref{FSnew}) from Model I, and tan$\beta/M_{H^\pm}=0.4$ GeV$^{-1}$\cite{MG2004, Aleph2001}.
In Figure 2, we include the uncertainty from the present experimental fit for $A_{\rm FB}$, which shows that the uncertainty could obscure the new physics signal, thus new physics searches through $A_{\rm FB}$ might be interesting in the future; however, they might be not very significant at present.

\begin{figure}[t]
\begin{center}
\includegraphics[width=12cm,height=10cm]{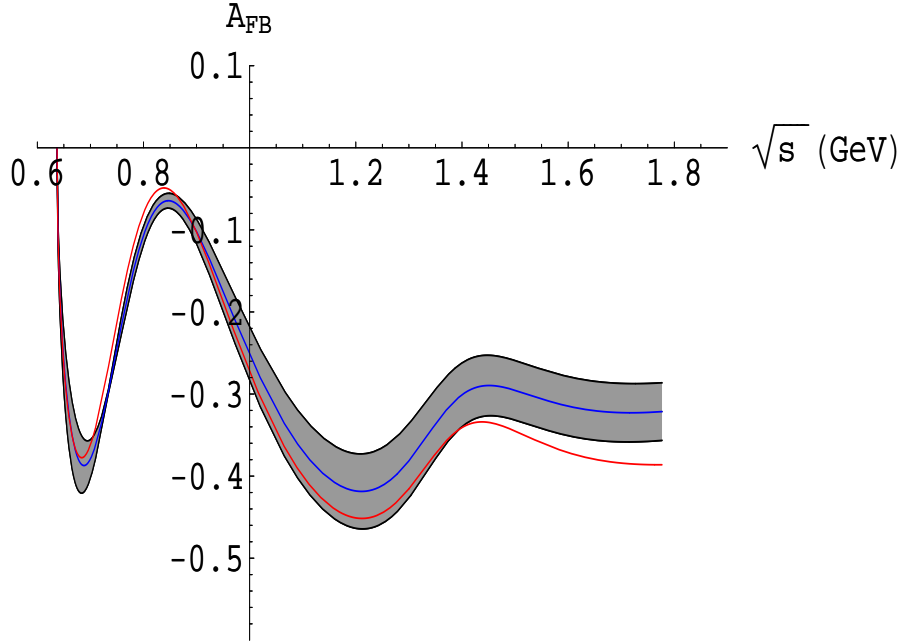}
\end{center}
\caption{The differential forward-backward asymmetry $A_{\rm FB}$ is plotted as the function of $\sqrt{s}$. The red line is from $F_S^{\rm New}$, the blue line is for Model I and grey region denotes the uncertainty from the parameters in Model I.}
\end{figure}
\section{Summary}

The Belle Collaboration reported a measurement of the decay spectrum of $\tau^- \to K_S\pi^-\nu_\tau$, and found two models, $K^*_0(800)+K^*(892)+K^*(1410)$ and $K^*_0(800)+K^*(892)+K^*_0(1430)$, that can both describe the spectrum very well. Based on these data, we carried  out a phenomenological analysis of the angular distributions in this decay.  We find that the $\cos\theta$-dependence of the normalized spectrum $1/\Gamma d\Gamma/d\cos\theta$, see eq.(\ref{angulardis2}), offers good opportunities to test the models. Therefore, the future experimental study of the angular distribution, in particular, the analysis of the forward-backward asymmetries, may be very useful to distinguish or rule out the above models, which would be helpful to reveal the nature of the scalar form factor of the decay. Possible new physics contributions from 2HDM to the $A_{\rm FB}$ have also been analyzed. We found that the present experimental uncertainty may obscure the new physics signal.

\vspace{0.5cm}
\section*{Acknowledgments}
This work was supported in part by the NSF of China under Grant Nos. 11075149 and 11235010, and the 973 project under Grant No. 2009CB825200.


\begin{thebibliography}{40}
\bibitem{Pich97}A. Pich,  {\it Tau physics}, Adv. Ser. Direct. High Energy Phys. {\bf 15}, 453 (1998)
 [hep-ph/9704453]; {\it Tau physics}, Int. J. Mod. Phys. A {\bf 15S1}, 157 (2000) [hep-ph/9912294]. A. Pich, I. Boyko, D. Dedovich, and I.I. Bigi, {\it Tau decays}, Int. J. Mod. Phys. A {\bf 24S1}, 715 (2009).
\bibitem{DHZ06}M. Davier, A. H\"ocker, and Z. Zhang, {\it The physics of hadronic tau decays}, Rev. Mod. Phys. {\bf 78}, 1043 (2006) [hep-ph/0507078].
\bibitem{Belle2011}M. Bischofberger {\it et al.}, Belle Collaboration, {\it Search for CP violation in $\tau^\pm \to K_S^0\pi^\pm\nu_\tau$ decays at Belle}, Phys. Rev. Lett. {\bf 107}, 131801 (2011), arXiv:1101.0349[hep-ex].
\bibitem{KM92}J.H. K\"uhn and E. Mirkes, {\it Strucuture functions in tau decays}, Z. Phys. C {\bf 56}, 661 (1992); Z. Phys. C {\bf 67}, 364 (1995), Erratum.
\bibitem{DM93}R. Decker, E. Mirkes, R. Sauer, and Z. Was, {\it Tau decays into three pseudoscalar mesons}, Z. Phys. C {\bf 58}, 445 (1993).
\bibitem{FM96}M. Finkemeier and E. Mirkes, {\it The scalar contribution to $\tau\to K\pi \nu_\tau$}, Z. Phys. C {\bf 72}, 619 (1996) [hep-ph/9601275].
\bibitem{ExpKstar}J.M. Dorfan {\it et al.}, MARK II Collaboration, {\it Measurement of the Branching Fraction for the Cabibbo-Suppressed Decay $\tau^-\to K^{*-}(892)\nu_\tau$}, Phys. Rev. Lett. {\bf 46}, 215 (1981); H. Albrecht {\it et al.}, {\it Measurement of the decays $\tau^-\to K^*\nu_\tau$} and $\tau^-\to\rho^-\nu_\tau$, ARGUS Collaboration, Z. Phys. C {\bf 41}, 1 (1988); M. Battle {\it et al.}, CLEO Collaboration, {\it Measurement of Cabibbo-Suppressed Decays of the $\tau$ Lepton}, Phys. Rev. Lett. {\bf 73}, 1079 (1994).
\bibitem{NC95}J.J. Godina Nava and G. Lopez Castro, {\it Tensor interactions and $\tau$ decays}, Phys. Rev. D {\bf 52}, 2850 (1995) [hep-ph/9506330].
\bibitem{Aleph99}R. Barate {\it et al.}, {\it One-prong $\tau$ decays with kaons}, ALEPH Collaboration, Eur. Phys. J. C {\bf 10}, 1 (1999) [hep-ex/9903014].
\bibitem{Cleo96}T.E. Coan {\it et al.}, {\it Decays of $\tau$ leptons to final states containing $K_S^0$ mesons}, CLEO Collaboartion, Phys. Rev. D {\bf 53}, 6037 (1996).
\bibitem{Belle2007}D. Epifanov {\it et al.}, {\it Study of $\tau^-\to K_S\pi^-\nu_\tau$ decay at Belle}, Belle Collaboration, Phys. Lett. B {\bf 654}, 65 (2007), arXiv:0706.2231[hep-ex].
\bibitem{BT95}L. Beldjoudi and T.N. Truong,{\it $\tau\to \pi K \nu$ decay and $\pi K$ scattering}, Phys. Lett. B {\bf 351}, 357 (1995) [hep-ph/9411423].
\bibitem{KLM12}D. Kimura, K.Y. Lee, and T. Morozumi, {\it The form factors of $\tau\to K\pi(\eta)\nu$ and the predictions of CP violation beyond the standard model}, arXiv:1201.1794 [hep-ph].
\bibitem{FZ99}S. Fajfer and J. Zupan, {\it The role of $K^*_0(1430)$ in $D\to PK$ and $\tau\to KP\nu_\tau$ decays}, Int. J. Mod. Phys. A {\bf 14},4161 (1999) [hep-ph/9903427].
\bibitem{JPP06}M. Jamin, A. Pich, and J. Portoles, {\it Spectral distribution for the decay $\tau\to \nu_\tau K\pi$}, Phys. Lett. B {\bf 640}, 176 (2006) [hep-ph/0605096]; {\it What can be learned from the Belle spectrum for the decay $\tau^-\to \nu_\tau K_S\pi^-$}, Phys. Lett. B {\bf 664}, 78 (2008), arXiv:0803.1786[hep-ph].
\bibitem{BEJ}D.R. Boito, R. Escribano, and M. Jamin, {\it $K\pi$ vector factor constrained by $\tau\to K\pi \nu_\tau$ and $K_{l_3}$ decays}, JHEP {\bf 1009}, 031 (2010), arXiv:1007.1858[hep-ph]; {\it $K\pi$ vector factor, dispersive constraints and $\tau\to \nu_\tau K\pi$ decays}, Eur. Phys. J. C {\bf 59}, 821(2009), arXiv:0807.4883[hep-ph].
\bibitem{KLMN09}D. Kimura, K.Y. Lee, T. Morozumi, and K. Nakagawa, {\it CP violation of $\tau\to K\pi(\eta,\eta^\prime)\nu$ decays}, arXiv:0808.0674[hep-ph]; D. Kimura, K. Nakagawa, T. Morozumi, and K.Y. Lee, {\it Direct CP violation in hadronic $\tau$ decays}, Nucl. Phys. B (Proc. Suppl. ) {\bf 189}, 84 (2009).
\bibitem{JOP}M. Jamin, J.A. Oller, and A. Pich, {\it S-wave $K\pi$ scattering in chiral perturbation theory with resonances}, Nucl. Phys. B {\bf 587}, 331 (2000) [hep-ph/0006045]; {\it Strangeness-changing scalar form factors}, Nucl. Phys. B {\bf 622}, 279 (2002) [hep-ph/0110193]; {\it Light quark masses from scalar sum rules}, Eur. Phys. J. C {\bf 24}, 237 (2002) [hep-ph/0110194]; {\it Scalar $K\pi$ form factor and light quark masses}, Phys. Rev. D {\bf 74}, 074009 (2006)[hep-ph/0605095].
\bibitem{2HDM}J.F. Gunion, H.E. Haber, G.L. Kane, and S. Dawson,
{\it The Higgs hunter's guide}, Report No. SCIPP-89-13.
\bibitem{MG2004}W.M. Morse, {\it Is the difference between the pion form-factor measured in $e^+ e^-$ annihilations and $\tau^-$ decays due to an $H^-$ propagator?}, hep-ph/0410062; D.N. Gao, {\it  Angular distribution asymmetry in $\tau^-\to\pi^-\pi^0 \nu_\tau$ decay in the two-Higgs-doublet model with large $\tan\beta$}, Phys. Rev. D {\bf 71}, 051301 [hep-ph/0411284].
\bibitem{Aleph2001}R. Barate {\it et. al.}, ALEPH Colloboration,{\it  Measurements of BR($b\to\tau^- \bar{\nu}_\tau X)$ and BR($b \to \tau^- \bar{\nu}_\tau D^{*+-} X)$ and upper limits on BR$(B^- \to \tau^- \bar{\nu}_\tau$) and BR $(b\to s \nu \bar{\nu})$}. Eur. Phys. J. C {\bf
19}, 213 (2001) [hep-ex/0010022].

\end{thebibliography}
\end{document}